\documentclass{aa}
\usepackage{psfig}
\usepackage{graphicx}

\newcommand{\be}{\begin{displaymath}}
\newcommand{\ee}{\end{displaymath}}

\newcommand{\et}{et al.}
\def\ergsec{erg\,s$^{-1}$}
\def\msun{M$_\odot$}
\def\ergcms{erg\,cm$^{-2}$\,s$^{-1}$}
\def\rwd{R$_{\rm wd}$}
\def\ros{ROSAT}
\def\xmm{XMM$-$Newton}
\def\dl{DP Leo}

\begin{document}

\title{An XMM-Newton timing analysis of the eclipsing polar DP Leo}

   \author{A.D.\,Schwope\inst{} \and
	   V.\,Hambaryan\inst{} \and
           R.\,Schwarz\inst{}}

   \offprints{A.D. Schwope, aschwope@aip.de}
   
   \institute{Astrophysikalisches Institut Potsdam,
              An der Sternwarte 16, D-14482 Potsdam, Germany}

   \date{Received; accepted}
   
\abstract{
We present an analysis of the X-ray light curves of the
magnetic cataclysmic variable DP Leo using recently 
performed \xmm\ EPIC and archival \ros\/ PSPC observations.
We determine the eclipse length at X-ray wavelengths to be $235\pm5$\,s, 
slightly longer than at ultra-violet wavelengths, where it lasts 225\,s.
The implied inclination and mass ratio for an assumed 0.6\,M$_\odot$
white dwarf are $i=79.7\degr$ and $Q = M_{\rm wd}/M_2 = 6.7$.
We determine a new linear X-ray eclipse and orbital ephemeris which 
connects the more than 120000 binary cycles covered since 1979. 
%Our periods deviate slightly from published values likely due to
%the former omission of leap seconds.  
Over the last twenty years, the optical and X-ray bright phases 
display a continuous shift
with respect to the eclipse center by $\sim 2.1\degr$\,yr$^{-1}$.
Over the last 8.5 years the shift of the X-ray bright phase 
is $\sim2.5\degr$\,yr$^{-1}$.
We interpret this as evidence of an asynchronously rotating
white dwarf although synchronization oscillations cannot 
be ruled out completely. If the observed phase shift 
continues, a fundamental rearrangement of the 
accretion geometry must occur on a time-scale of some ten years.
DP Leo is marginally detected at eclipse phase.
The upper limit eclipse flux is consistent with an origin 
on the late-type secondary, 
$L_{\rm X} \simeq 2.5 \times 10^{29}~{\rm ergs~s}^{-1} 
(0.20-7.55\, {\rm keV})$,
at a distance of 400 pc.
}
%\authorrunning{Schwope et al.}
%\titlerunning{}
  
\maketitle

\section{Introduction}\label{Intro}
DP Leo is one of the strongly magnetic cataclysmic binaries of 
AM Her type, a so called polar. It was discovered as the first 
eclipsing polar some 20 years ago
as optical counterpart of the EINSTEIN source E1114+182 (Biermann et al.~1985),
and continuously observed with from the ground and
the space with e.g.~HST (Stockman et al.~1994) and ROSAT 
(Robinson \& Cordova 1994, hereafter RC94). 
It was found to be a two-pole accretor 
based on the detection of cyclotron emission lines in field strengths
of 30.5\,MG and 59\,MG, respectively (Cropper \& Wickramasinghe 1993). 
A thorough timing study by Robinson \& Cordova (1994) using ROSAT X-ray data 
combined with earlier optical data revealed evidence for an asynchronous 
rotation of the white dwarf in the system. 
Asynchronous polars form a very small subgroup of all polars. There are four
out of currently known 65 systems which show a small degree of asynchronism 
of typically
about 1\% (Campbell \& Schwope 1999). With an extra spin of the white 
dwarf in DP Leo of about 2\degr--2.5\degr\ per year, 
the degree of asynchronism 
is seemingly much smaller than in the other four objects. However, the 
earlier results are based on a mixture of optical and X-ray data 
with not necessarily common origin on the white dwarf. 

DP Leo was chosen as Calibration/Performance Verification 
target of XMM-Newton and was observed with all three X-ray 
telescopes in Nov.~2000. The spectrum derived from these observations
was recently published by Ramsay et al.~(2001). Using a multi-temperature 
model of the post-shock flow, they found evidence 
of a very massive white dwarf in excess of 1\,M$_\odot$. 
In order to address the question of asynchronism in DP Leo based 
on X-ray data alone, we performed a timing analysis of the new 
XMM-data in combination with archival ROSAT observations (one published 
by Robinson \& Cordova, a second one unpublished, Sect.~\ref{s:dat}). 
For proper measurement of the eclipse parameters we used segmented
data, where individual segments were determined with
a Bayesian change point detection method (Sect.~\ref{s:bay}).
Our main results are presented in Sect.~\ref{s:res}, where the eclipse
parameters, an updated eclipse ephemeris and the accretion geometry
are discussed.
The question whether there is a positive detection of the secondary 
at X-ray wavelengths in the eclipse is discussed in Sect.~\ref{s:sec}.

\section{Observations and Data reduction} \label{s:dat}

\subsection{\xmm\ EPIC}
DP Leo was observed using \xmm\ on 22 of November 2000 for a 
net exposure time of 19949~s. 
DP Leo was detected in all three EPIC detectors
(Turner et al. \cite{turner}, Str\"uder et al. \cite{struder}). 
The thin filter was used and the CCDs were read out 
in full window mode. 

Before extracting source photons, the data were processed using the
current release of the \xmm\ Science Analysis System (version 5.1).
Standard procedures of data screening (creation of an image and a background 
light curve) revealed time intervals with enhanced particle background. 
These intervals were excluded from the subsequent 
timing analysis using an approach 
described below (See Sect.~\ref{s:bay}). This reduces the 
accepted exposure time to 
15827~sec with the EPIC-PN detector. The observations were 
performed without any interruption, i.e.~full phase-coverage 
of the $P_{\rm orb} = 5388$\,s binary was achieved with an average 
exposure of $\sim$150\,s per 0.01 phase unit. 
%The time resolution 
%of the EPIC PN is 73 msec, that of the EPIC-MOS cameras 2.6\,s, the
%latter being too low for detailed timing studies.

\begin{figure}[t]
\resizebox{\hsize}{!}{\includegraphics[bbllx=54pt,bblly=211pt,bburx=460pt,bbury=673pt,clip=]{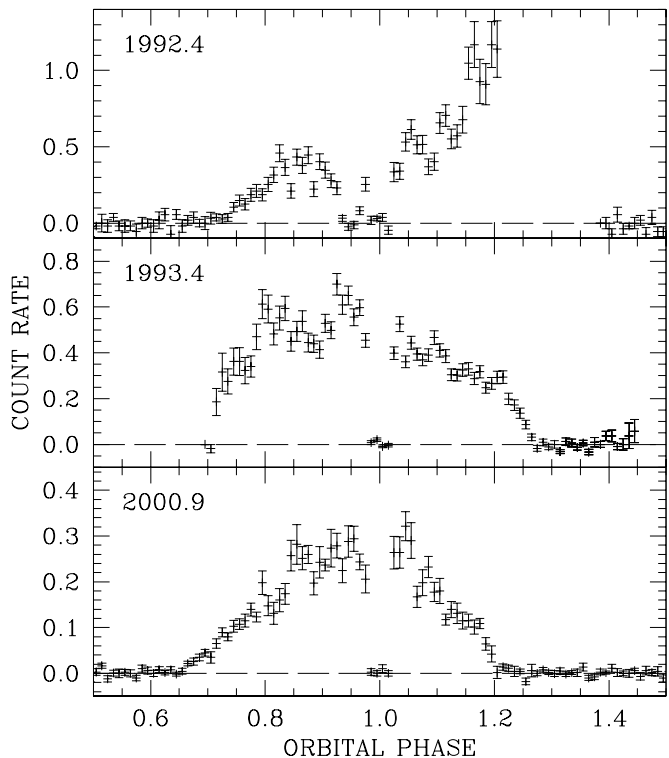}}
\caption{Phase-averaged X-ray light curves of the \ros\ 
and the \xmm\ observations. Phase bins have a size of
0.01 phase units of the $P_{\rm orb} = 5388$\,s binary.
}
\label{f:xlcs}
\end{figure}

\subsection{\ros\ PSPC }

The field of DP Leo was also observed with the \ros\ PSPC on 
May 30, 1992 (ROR 300169, PI: Cordova) and on May 30, 1993 
(ROR 600263, PI: Petre) for a net exposure time of 8580~s and 23169~s, 
correspondingly. Results of the 1992 observations were presented by 
Robinson \& Cordova (1994), the results of the much more extended 
observations of 1993 are unpublished. 

Although the net exposure time of the two \ros\ observations was larger
than the binary period, in
{\bf %rev
neither case was complete phase coverage 
achieved due to the close 
proximity of the periods of the satellite and of the binary. }%rev

There are further X-ray observations reported by 
Biermann et al.~(1985), and Schaaf et al.~(1987), respectively, 
with the EINSTEIN and EXOSAT satellites. 
We make use of the timing of the X-ray eclipses detected with 
Einstein, the EXOSAT data have a too low count-rate and are not 
used further in this paper.

%\begin{figure}[t]
%\resizebox{\hsize}{!}{\includegraphics[bbllx=-12pt,bblly=-4pt,bburx=628pt,bbury=743pt,angle=-90,clip=]{mod_specs.ps}}
%\caption{Modeled spectra of the \ros\ (1993) and the \xmm\ (2000) observations
%of \dl.
%}
%\label{f:spec}
%\end{figure}

\subsection{Timing analysis with a Bayesian change point detection method}
\label{s:bay}

In order to study the abrupt changes of the X-ray count rate 
particularly at eclipse ingress and egress we performed a 
timing analysis of the datasets using a Bayesian change point 
detection method\footnote{In general terms, 
the change-point methodology deals with sets of 
sequentially ordered observations (as in time) and determines
whether the fundamental mechanism generating the observations has changed
during the time the data have been gathered (see, e.g. Csorg\"o \& 
Horv\'ath, \cite{chp}).} developed by Scargle (\cite{scar1,scar2}). 
This method is well suited for a  
statistical examination when the arrival times of individual 
X-ray photons are registered (see Hambaryan \et\ \cite{me}).
It is superior to methods which work on binned data, since it requires 
no a priori knowledge of the relevant time-scale of the structure 
which will be investigated.

The method is applicable to data that are known to originate from  
a nearly ideal Poisson process, i.e. a class of independent, identically
distributed processes, having zero lengths of dead time.
The data gathered in \xmm\ EPIC-PN  and \ros\ PSPC observations 
allow the measurement of arrival times of individual 
X-ray photons with a resolution of 73.3~ms and 
0.1~ms, respectively,
a resolution much smaller than the ingress and egress time scale which 
is of the order of seconds. The EPIC-MOS data cannot be used for the
study of the eclipse length, since it provides a resolution of 
only 2.6\,s.

Scargle's (\cite{scar1,scar2}) method 
decomposes a given set of photon counting data into 
Bayesian blocks with piecewise constant count-rate according to 
Poisson statistics.
Bayesian blocks are built by a Cell Coalescence algorithm 
(Scargle \cite{scar2}), which begins  with a fine-grained segmentation.
It uses a Voronoi tessellation\footnote{The Voronoi cell for a 
data point consists of all the space closer to that point than to
any other data point.} of data points, where neighboring cells
are merged if allowed by the corresponding marginal likelihoods 
(see Scargle \cite{scar2}).

{\bf %rev
We repeat here the essential parts of the method, expanding upon particular
modifications of the original method as used in the present application.
Assume that during a continuous
observational interval of length $T$, consisting of $m$ discrete
moments in time (spacecraft's ``clock tick''),  a set of
photon arrival times $D$ ($t_i, t_{i+1},...,t_{i+n}$) is registered. 
Suppose now that we want to use these data to
compare two competing hypotheses, The first hypothesis is that the
data are generated from a 
constant rate Poisson process (model $ M_1$) 
and the second one from two-rate Poisson
process (model $ M_2$). Evidently, model $ M_1$ is described by only 
one parameter $\theta$ (the count rate) 
of the one rate Poisson process while the model 
$ M_2$ is described by
parameters $\theta_1$, $\theta_2$ and $\tau$. The parameter $\tau$
is the time when the Poisson process switches from
$\theta_1$ to $\theta_2$ during the total time $T$ of observation, 
which thus is divided in intervals $T_1$ and $T_2$.

By taking as a background information ($I$) the proposition that one
of the models under consideration is true and by using Bayes' theorem
we can calculate the posterior probability  
of each model by (the probability that $M_k$
($k=1,2$) is the correct model, see, e.g., Jaynes \cite{jaynes})
\begin{equation}
Pr(M_k|D,I)=\frac{Pr(D|M_k,I)}{Pr(D|I)}Pr(M_k|I)
\label{mek}
\end{equation}
where $Pr(D|M_k,I)$ is the (marginal) probability of the data assuming model
$ M_k$, and $Pr(M_k|I)$ is the prior probability of model $ M_k\; (k=1,
2)$. The term in the denominator is
a normalization constant, and we may eliminate it by calculating the ratio
of the posterior probabilities instead of the probabilities directly. 
Indeed, the extent to which the data support model $ M_2$
over $ M_1$ is measured by the ratio of their posterior probabilities 
and is called the posterior odds ratio

\begin{equation}
O_{21}\equiv\frac{Pr(M_2|D,I)}{Pr(M_1|D,I)}=\left[\frac{Pr(D|M_2,I)}{Pr(D|M_1,I)}\right]\left[\frac{Pr(M_2|I)}{Pr(M_1|I)}\right].
\label{erku}
\end{equation}

The first factor on the right-hand side of Eq. (2) is the ratio
of the \emph{integrated} or \emph{global} likelihoods of the two
models and is called the
\emph{Bayes factor} for $ M_2$ against $ M_1$, denoted by $B_{21}$. The
global likelihood for each model can be evaluated by integrating over
nuisance parameters and the final result for discrete Poisson events
can be represented by (see, for details, Scargle \cite{scar1,scar2}, 
Hambaryan et al.~1999)

\begin{eqnarray}
B_{21}=\frac{1}{B(n+1,m-n+1)}\sum{B(n_1+1,m_1-n_1+1)}
\nonumber \\ \\
\times B(n_2+1,m_2-n_2+1)\Delta \tau \,, \nonumber
\label{erek}
\end{eqnarray}
where $B$ is the \emph{beta function}, $n_j$ and $m_j, (j=1,2)$, respectively are the number of recorded
photons and the number of ``clock ticks'' in the observation
intervals of lengths $T_1$ and $T_2$. $\Delta \tau$ is the time
interval between successive photons, and the sum is over the photons' index.

\begin{figure}[t]
%\resizebox{\hsize}{!}{\includegraphics%
%[bbllx=68pt,bblly=38pt,bburx=549pt,bbury=698pt,clip=]
%{recipro_lc.ps}}
%\resizebox{\hsize}{!}{\includegraphics%
%[bbllx=49pt,bblly=190pt,bburx=563pt,bbury=497pt,clip=]
%{pn_ecl_lc_5s.ps}}
\resizebox{\hsize}{!}{\includegraphics%
[bbllx=49pt,bblly=189pt,bburx=457pt,bbury=707pt,clip=]
{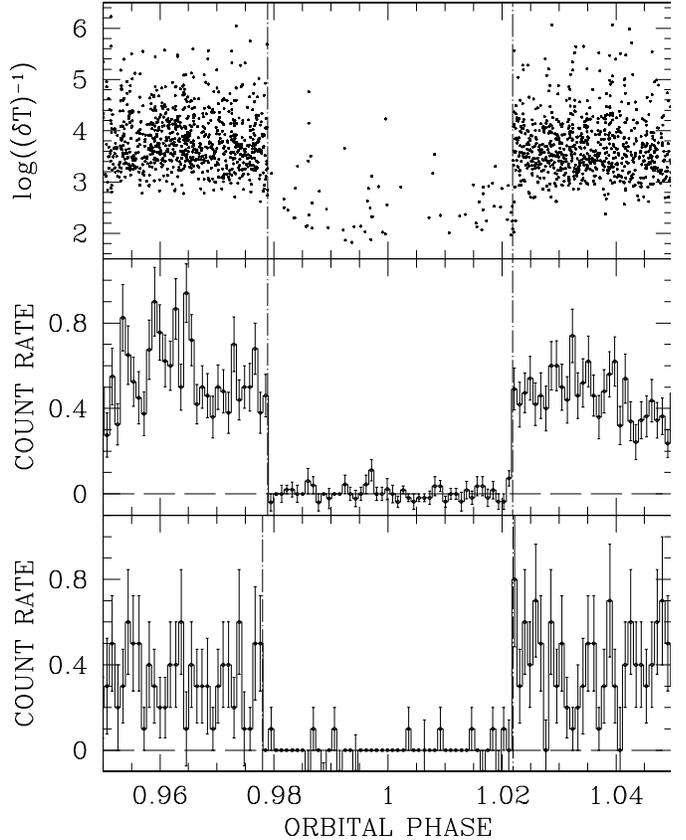}}
\caption{Determination of the eclipse length for the \ros\ 1993 
(upper two panels) and the \xmm\ observations (lower panel) 
of DP Leo. The upper panel shows the distribution 
of the reciprocals of the time intervals between neighbouring 
photons, the two lower panels binned X-ray light curves (bin size 5 sec). 
Vertical lines indicate the change points in the Poisson process, the 
space between them is our measured eclipse length. 
}
\label{f:reci}
\end{figure}

The second factor on the right-hand side of Eq. (\ref{erku}) is
the prior odds ratio, which
will often be equal to 1 (see below), representing the absence
of an {\it a priori} preference for either model.

It follows that the Bayes factor is equal to the posterior odds when the
prior odds is equal to 1. When $B_{21}\;>\;1$, the data favor $ M_2$
over $ M_1$, and when $B_{21}\;<\;1$ the data favor $ M_1$.

If we have calculated the odds ratio $O_{21}$, in favor of model $ M_2$
over $ M_1$, we can find the probability for model $ M_2$ by inverting
Eq.~(\ref{erku}), giving

\begin{equation}
Pr(M_2|D,I)=\frac{O_{21}}{1+O_{21}} \,.
\label{chors}
\end{equation}

Applying this approach to the observational data set,
Scargle's (\cite{scar1,scar2}) method returns an array of rates,
$(\theta_1, \theta_2,..., \theta_{cp})$, and a set of so called ``change
points'' $(\tau_1, \tau_2,...., \tau_{cp-1})$, giving the times when an
abrupt change in the rate is determined, i.e. a significant variation. This is the most probable partitioning of the observational interval into blocks
during which the photon arrival rate displayed no 
statistically significant variations.
} %rev

We determined the timing accuracy of these change points 
through simulations. 
We generated 1000 data sets (photon arrival times) 
with one change point each. The data in the two segments
obeyed Poisson statistics.
Each simulated data set had approximately 
the same characteristics as
the observed data in terms of number of registered counts, spanned time, 
characteristic time scales of expected variations, 
and was analyzed exactly in the same way. 
The standard deviation of the distribution of change points was found to 
be $\Delta t_{cp} = \pm (2-3)$~s, if phase-folded data are used. 
The uncertainty was larger, when data in original time sequence were used
due to the smaller total number of photons involved.
We adopted an uncertainty of 2.5\,s for the observationally 
determined change points which were used to derive the eclipse 
length.

{\bf %rev
In Fig.~\ref{f:reci} we visualize the outcome of the process for the
\ros\ observations performed in 1993 and for the XMM-Newton observations
(EPIC-PN data only). The top panel shows the 
distribution of the reciprocal of the time interval between neighbouring 
photons (1993 data), in the two lower panels the X-ray light curves 
binned in intervals 
of 5 sec are plotted. The change points determined
by our method are indicated by vertical lines.
}%rev

\begin{table*}
\caption{Features of the X-ray light curve derived from \ros\ and \xmm\
observations of \dl. For a given epoch and instrument we list 
the mean count rate in the phase interval 0.80-0.90, 
the length of the eclipse $\Delta t_{\rm ecl}$,
the phase of the center of the bright phase interval $\phi_{\rm C}$,
and the length of the bright phase $\Delta \phi_{\rm B}$.
{\bf %rev
The center of the bright phase is interpreted as accretion spot longitude.
}%\rev
\label{t:feat}
}
\begin{tabular}{llcccc}\hline
Epoch  &  Mission/Det &      CR & %$T_0$ & 
$\Delta t_{\rm ecl}$ & 
$\phi_{\rm C}$ & 
$\Delta \phi_{\rm B}$\\
& & [s$^{-1}$]% & [MJD] 
& [s] & &\\
\hline
1992.4 & ROSAT/PSPC & $0.35$ & %2448773.215107(52) & 
	$237\pm5$ & $0.006\pm 0.006$& --\\%    +02

1993.4 & ROSAT/PSPC & $0.50$ & %2449137.913051(52) & 
	$233\pm5$ & $0.013\pm 0.006$& 0.57\\%     +4.6+/-2

2000.9 & \xmm/EPIC & $0.25$ & %2451870.776884(52) & 
	$237\pm5$ & $0.067\pm 0.006$& 0.57\\% +24+/-2
\hline
\end{tabular}
\end{table*}

\subsubsection{\xmm\ EPIC PN}\label{pn}

As a first step, we applied the change point detection 
method to a background region free of any X-ray source. 
This allowed us to determine time intervals where the
background showed no significant variation. These were regarded
as good time intervals and further used for the timing analysis of DP Leo. 

We extracted $\sim 3180$ EPIC-PN  photon events from the source, 
whose arrival times were corrected to the solar system barycenter 
using the ``barycen'' task, as implemented in SAS version 5.1. 

We used phase-folded data and data in original time sequence in 
order to determine different quantities. The length of the bright
phase and the eclipse length were measured in phase-folded data,
the times of individual eclipses (for a period update) 
were measured in original time sequence.
%We performed a change point detection search on phase-folded 
%date, i.e.~in phase space, using all photons in the good time
%intervals in order to determine the eclipse length at the given epoch. 
%For phase-folding we used the period given by RC94.

The mean bright-phase count-rate, the eclipse length, bright phase center 
and length of bright phase are listed in Tab.~\ref{t:feat}, whereas the
times of mid-eclipse of the individual eclipses are listed in 
in Tab.~\ref{t:ecl}. The times given there are barycentric 
Julian ephemeris days, i.e.~they take into account 
the 14 leap-seconds introduced between the first and the last 
data point. Since leap-seconds were omitted by seemingly all authors
in the past 
{\bf%rev
(all timings in the literature are given in HJD only),}%rev
we computed for all eclipse times we found in the literature
the leap-second correction and include those times 
in the table  for consistency and future work.

\subsubsection{\ros\ PSPC}
We also performed a timing analysis of the \ros\ PSPC observations.
In both cases we used a  circular region with $30\arcsec$ radius 
to extract source photons. In total 2705 and 8385 source 
counts were extracted, respectively. 
The radius chosen encompasses 85\% of the events in 
the \ros\ point spread function. Photon arrival times were
corrected to the solar barycenter, as implemented in 
the Extended Scientific  Analysis Software System 
(Zimmerman et al. \cite{zimm}). 

These data were treated in the same manner as the \xmm\ data and the 
corresponding results are also listed in Tabs.~\ref{t:feat} and \ref{t:ecl}.

\section{Results and discussion}\label{s:res}

\subsection{X-ray light curves and mean spectra of DP Leo}
The phase-averaged X-ray light curves of the two \ros\ and the 
\xmm\ observations (summed signal from all three cameras) are shown
in Fig.~\ref{f:xlcs}. At all occasions the source showed a 
pronounced on/off behavior with the eclipse roughly centered
on the X-ray bright phase. The eclipse was covered 3 times in 1992, 
8 times in 1993, and 2 times in the PN-observation (good time
intervals only, one eclipse was excluded from the analysis due to high 
particle background). 

In the 1992 observation the source showed a pronounced flare
at phase 0.2. The end of the bright phase was not covered, the length
of the bright phase, however, was inferred by RC94 from contemporaneous 
optical photometry.
A pre-eclipse dip, likely due to the intervening accretion stream
occurred centered at phase 0.94. Interestingly, this feature was never
observed again, indicating a re-arrangement of the accretion geometry.

The 1993 observation covered the X-ray bright phase completely
(although marginally at the start)
thus allowing to measure the length of the bright phase from X-ray data
alone. The source displayed similar brightness during the two \ros\
observations. The eclipse appeared centered on the bright phase.

In 2000, the shape of the X-ray bright phase appeared almost unchanged
compared to the 1993 observation. 
The eclipse now was clearly off-centered with respect to the bright 
phase. The rise to the bright phase 
was somewhat less steep than the fall. Compared with the earlier \ros\ 
observations, DP Leo appeared fainter in the center of the bright phase.
According to Ramsay et al.~(2001) and Pandel et al.~(2001)
\dl\ was in a state of intermediate accretion at the time 
of the \xmm\ observations, whereas it was in a high state 
at the time of the \ros\ observations. The comparison of 
published results combined with our own analysis 
shows that the situation might be different.

For the PSPC observations of 1992, RC94 derive a bolometric 
blackbody luminosity for an assumed distance of 260\,pc of
$L_{\rm bb, bol} = \kappa \pi F_{\rm bb} = 1.4^{+7.1}_{-0.3}
 \times 10^{31}$\,\ergsec. Scaling to the more likely distance of
400\,pc gives $L_{\rm bb, bol} = 3.3 \times 10^{31}$\,\ergsec. 
RC94 used a geometry factor $\kappa = 2$. Ramsay et al.~(2001) used
$\kappa = \sec(i - \beta) = \sec(80\degr - 100\degr)= 1.06$ 
and a distance of 400\,pc and derive 
$1.5\times 10^{31}$\,\ergsec with the EPIC MOS detectors, more than twice 
that value with the EPIC PN detector. 
Within the accuracy of the measurements and scaled to the same geometry
factors the luminosities of the soft components at both epochs 
agree with each other.

Contrary to the PSPC observations in 1992, there is a clear 
detection of DP Leo above 0.5\,keV in the PSPC observation 
performed in 1993, which allows fitting 
of a two-component spectrum. With the spectral resolution 
provided by the  ROSAT PSPC, the spectrum
is well reflected by a combination of a black-body  and a bremsstrahlung 
component. We fixed the bremsstrahlung temperature at 
the typical temperature of $kT_{\rm br} = 15$\,keV. The bolometric
flux in the bremsstrahlung component thus derived was 
$F_{\rm br, 93} = 2.9 \times 10^{-13}$\,\ergcms. Application 
of the same simple model to the EPIC PN data, and adding 
a Gaussian for the iron line at 6.7\,keV, gives a fitted temperature of  
$kT_{\rm br} = 11\pm6$\,keV and a bolometric flux of 
$F_{\rm br, 00} = 2.4 \times 10^{-13}$\,\ergcms, which again is
not in contradiction to the former \ros\ measurements. We conclude
that the X-ray observations do not indicate an
obvious change of the mass accretion rate between the three epochs.

\begin{table}
\caption{Times of mid-eclipse of all eclipses measured including new \ros\ and 
\xmm\ data. Individual times are leap-second corrected times 
at the solar system barycenter (BJED: barycentric Julian ephemeris day).
\label{t:ecl}
}
\begin{tabular}{lrlrc}
      \hline
      \noalign{\smallskip}
Epoch  &  Cycle & BJED   & $\delta T$ & Type$^{(1)}$ \\
       &        & --2400000 &   (sec)  &              \\
\hline
1979.9 & $-$73099 & 44214.55325& 15 & X \\
1979.9 & $-$73098 & 44214.61562& 15 & X \\
1979.9 & $-$73097 & 44214.67798& 15 & X \\
1982.0 & $-$61017 & 44968.02309&100 & O \\
1982.0 & $-$61002 & 44968.95712&100 & O \\
1982.0 & $-$61001 & 44969.01962&100 & O \\
1982.0 & $-$60841 & 44978.99755&100 & O \\
1982.1 & $-$60602 & 44993.90078& 60 & O \\
1982.1 & $-$60601 & 44993.96328& 60 & O \\
1982.1 & $-$60600 & 44994.02642& 60 & O \\
1982.1 & $-$60169 & 45020.90513& 20 & O \\
1982.1 & $-$60153 & 45021.90292& 20 & O \\
1982.2 & $-$60106 & 45024.83386& 60 & O \\
1984.1 & $-$48767 & 45731.96640& 30 & O \\
1984.2 & $-$48256 & 45763.83373&  5 & O \\
1984.4 & $-$46796 & 45854.88280&100 & X \\
1985.0 & $-$43588 & 46054.94231&100 & X \\
1985.1 & $-$43075 & 46086.93565&  3 & O \\
1985.1 & $-$43074 & 46086.99796&  3 & O \\
1991.8 &  $-$3410 & 48560.55789&  4 & UV \\
1992.4 &      0 & 48773.21509&  5 & X \\
1992.4 &     16 & 48774.21293&  5 & X \\
1993.4 &   5848 & 49137.91294&  5 & X \\
1993.4 &   5945 & 49143.96214&  5 & X \\
1993.4 &   5946 & 49144.02438&  5 & X \\
1993.4 &   5947 & 49144.08689&  5 & X \\
1993.4 &   5961 & 49144.96005&  5 & X \\
1993.4 &   5962 & 49145.02235&  5 & X \\
1993.4 &   5963 & 49145.08454&  5 & X \\
1993.4 &   5964 & 49145.14711&  5 & X \\
2000.9 &  49670 & 51870.77761&  5 & X \\
2000.9 &  49672 & 51870.90237&  5 & X \\
 \noalign{\smallskip}
\hline
 \noalign{\smallskip}
\end{tabular}\\
   \noindent{\noindent\footnotesize$^{(1)}$ X~=~X-ray; O~=~optical; UV~=~UV}
\end{table}

%In our study we concentrate on the determination of 
%the length of the eclipse at X-ray wavelengths, its absolute timing, 
%and the centering of the bright phase 
%with respect to the eclipse. While the latter can be read using 
%cursor from a screen, the former need some sophistication. We 
%use for our study the Bayesian change point detection described above.

\begin{figure}[t]
\resizebox{\hsize}{!}{\includegraphics%
%[bbllx=54pt,bblly=211pt,bburx=557pt,bbury=673pt,clip=]
{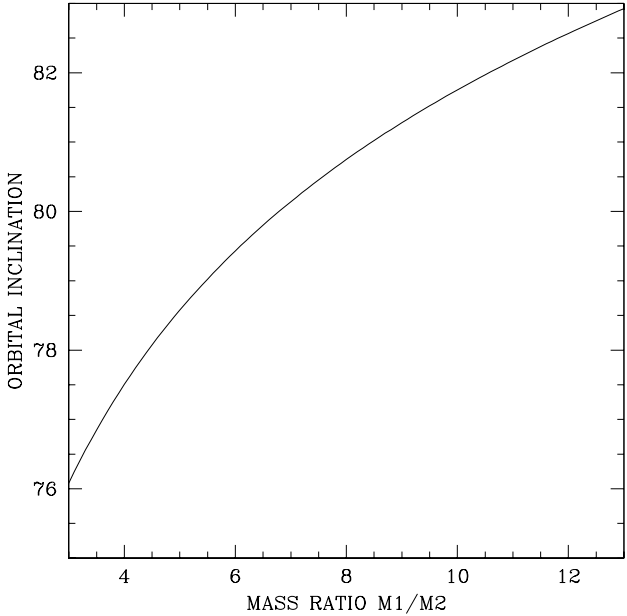}}
\caption{Relation between mass ratio and inclination for an 
eclipse length of 225\,s.
}
\label{f:qi}
\end{figure}

\subsection{Timing of the X-ray eclipse}
Application of our method to the data of \dl\ allowed an accurate
determination of the 
eclipse length at X-ray wavelengths.
The measurements at all three 
epochs agree with each other within the claimed accuracy, 
{\bf%rev 
ranging 
from 233\,s to 237\,s with a 5 second accuracy (see Tab.~\ref{t:feat}). 
}%rev
The average 
eclipse length is somewhat longer than that deduced from HST/FOS
observations. Figure 5 of the paper by Stockman et al.~(1994) implies
that the eclipse length at ultraviolet wavelengths is about 225\,s
(measured at half intensity).
The difference in the length of the ultraviolet and X-ray 
eclipses is due to the fact, that the source of X-ray emission 
is closer to the secondary star than the source of the ultraviolet 
radiation. The former originates from the hot accretion spot while 
the latter has contributions from the whole surface 
of the white dwarf. We regard the eclipse length determined in 
ultraviolet data as relevant for the mass determination.
Using the relation between eclipse length
(in phase units), inclination and mass ratio given by Chanan et al.~(1976)
and a eclipse length of 0.0209 phase units (225\,s), the relation between 
mass ratio and inclination as shown in Fig.~\ref{f:qi} results.

At the given period, the mass-radius relation for late-type stars by 
Caillault \& Patterson (1990) predicts a secondary star with 
mass of only 0.09\,M$_\odot$. Assuming a typical white dwarf with 
$M_{\rm wd} = 0.6\,M_\odot$, 
the mass ratio is $Q = M_{\rm wd} / M_2 = 6.7$ 
and the implied orbital inclination
$i = 80\degr$ (cf.~Bailey et al.~1993). 
Ramsay et al.~(2001) argued for a high mass white dwarf near the
Chandrasekhar limit on the
basis of their spectral model applied to the \xmm\ data. 
This would imply a slightly higher inclination 
of $i > 82\degr$. 
{\bf%rev
However, the comparison of white dwarf masses based on their 
model with dynamically determined masses in the well-studied 
polars QQ Vul 
($M_{\rm dyn} = 0.54$\,\msun vs.~$M_{\rm Xfit} = 1.30$\,\msun; 
Catal\'an et al.~1999, Cropper et al.~1999) 
and AM Her 
($M_{\rm dyn} = 0.45$\,\msun vs.~$M_{\rm Xfit} = 0.74$\,\msun; 
Schwarz et al.~2001, Ramsay et al.~2000) 
shows that the X-ray spectral}%rev 
model tends to predict a too high mass for the white dwarf. A high mass
white dwarf seems unlikely to us given the good fit shown by Bailey
et al.~(1993) to their optical eclipse light curve. A massive white 
dwarf would have less than half the radius of the 0.7\,M$_\odot$ 
white dwarf used by Bailey et al.~and would not give a comparably 
good fit. We use, therefore,
as a baseline for our further analysis the standard value 
$M_{\rm wd} = 0.6\,M_\odot$.

The new determination of the eclipse length has a much higher accuracy 
than that by RC94, who give $216\pm18$\,s and demonstrates the 
benefit of the Bayesian change point method.
RC94 derive an upper 
limit of $\sim$22\,s on the length of the eclipse ingress/egress
phase. 
%We tried to improve on this and performed extensive simulations.
%We generated photon lists with 
%a gradual increase/decrease of the count rate at ingress/egress
%phase lasting 5 seconds. We then searched for Bayesian change points and
%found only in about 10\% of all cases more than one 
%change point. This means, if the ingress/egress phase lasts shorter than
%about 5\,s, 
%it cannot be resolved with the present data. If ingress/egress phase would 
%have lasted longer, we would have a fair chance to detect this.
%Compared to the claimed 
%upper limit of $\sim$22\,s for the duration of ingress/egress 
{\bf %rev
The binned eclipse light curves of Fig.~\ref{f:reci}
clearly show, that eclispe ingress and egress lasts much shorter
than 22 sec in the observations perfomed in 1993 and 2000
but the count rate is not sufficient to resolve ingress and egress.
We therefore cannot derive strong  
constraints on the lateral extent of the X-ray emission 
region. For comparison, in UZ For and HU Aqr where the egress phases 
could be resolved by EUVE and ROSAT observations, respectively,
these features last only about 1.3\,s (Warren et al.~1995,
Schwope et al.~2001a), corresponding to a full opening angle of the 
X-ray emission region on the white dwarf of only 3\degr.
}%rev

We proceed by updating the eclipse ephemeris of DP Leo by using 
X-ray data alone. We disregard optical data for this purpose 
since it is shown for other polars (e.g.~in HU Aqr, Schwope et al.~2001b),
that the optical and X-ray emitting regions might be disjunct. 
This could result in a shift of the X-ray with respect to the optical
eclipse by several seconds and would corrupt the period 
determination. A linear regression to the eclipse times measured 
with EINSTEIN, ROSAT, and \xmm\ yields 
%\begin{equation}
\be
\mbox{BJED}_{\rm X,ecl} = T_{\rm 0,ecl} + E \times P_{\rm ecl}
\ee
\be
\mbox{BJED}_{\rm X,ecl} = 2448773.21503(2) + E \times 0.0623628471(5)
\ee
%\end{equation}
for the barycentrically and leap-second corrected time of the 
X-ray eclipse center. This is not to be mixed up with the orbital 
period of the binary system, since we are measuring the eclipse 
of a small structure on the white dwarf surface which obviously is not 
fixed in the binary system. Our determination of $P_{\rm ecl}$
is consistent with that of RC94 only at the 2.5$\sigma$ level with 
our period being longer. One reason
for the slight inconsistency could be the omission of leap seconds 
by RC94, another the use of different types of input data,
X-ray and optical data by RC94, X-ray data alone by us.
 
\subsection{Spot geometry and true phase zero}
The derivation of the true binary period of the system
needs independent information about the eclipse of the white
dwarf center (not the spot on it!).
There is one direct measurement in the literature (Stockman et al.~1994)
available to us based on HST/FOS measurements.
One can correct from the observed X-ray mid-eclipse times 
to the times of mid-eclipse of the white dwarf. We did this for the
eclipse data which entered the determination of $P_{\rm ecl}$ above.
This correction is based on the following parameters:
$i = 79.65\degr, Q = 6.7, M_{\rm wd} = 0.6$, 
mass-radius relations for the white dwarf and the secondary by
Nauenberg (1972) and Caillault \& Patterson (1990), 
spot latitude 100\degr,
spot longitude at the different epochs as listed in Tab.~\ref{t:feat},
the longitude at the time of the EINSTEIN observation was --22\degr.
Usage of these parameters gives the correct eclipse length and length 
of the bright phase, if a height of the emission region of 0.02\,\rwd\
is taken into account. The assumed height is in accord with other 
polars (e.g.~Schwope et al.~2001a).
The small corrections to the eclipse times as listed in the above 
table are $+5.7, -0.5, -1.2, -6.2$\,s for the EINSTEIN, \ros\ 1992, 
\ros\ 1993, and the \xmm\ observations, respectively. 
A linear regression to the corrected timings including the HST
measurement gives the ephemeris of inferior conjunction of the 
secondary 
\be
\mbox{BJED}_{\rm XUV,orb} = T_{\rm 0,orb} + E \times P_{\rm orb}
\ee
\be
\mbox{BJED}_{\rm XUV,orb} = 2448773.21503(2) + E \times 0.0623628460(6).
\ee
The binary period $P_{\rm orb}$ is slightly shorter than $P_{\rm ecl}$, 
as expected. Both zero points, $T_{\rm 0,orb}$ and 
$T_{\rm 0,ecl}$, agree with each other, since the spot was at about 
longitude zero at the epoch, when our cycle counting starts.
The introduction of a quadratic term, as  discussed by RC94,
gives a bad fit to the data with an $(O - C)$ time of about one minute at the 
epoch of the \xmm\ observations and is therefore ruled out.

\begin{figure}[t]
\resizebox{\hsize}{!}{\includegraphics%
%[bbllx=54pt,bblly=211pt,bburx=557pt,bbury=673pt,clip=]
%{shift_bripha.ps}}
%[bbllx=70pt,bblly=540pt,bburx=720pt,bbury=62pt,angle=90,clip=]%
[angle=90]%
{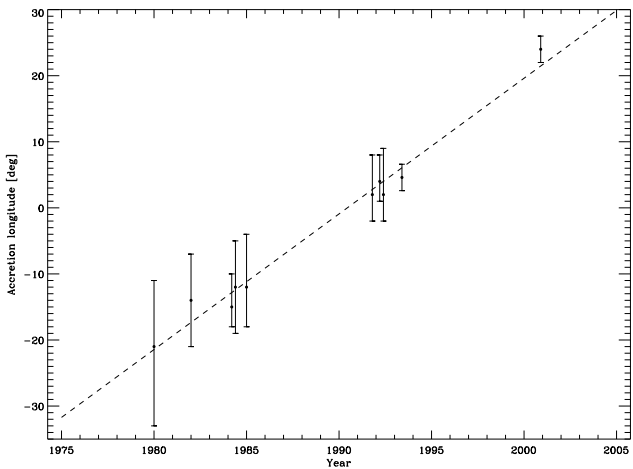}}
\caption{Accretion spot longitude variations as a function 
of time. The last two data points are from this work, the other 
points are from RC94. The dashed line is a linear fit to all data.
}
\label{f:bripha}
\end{figure}

\subsection{Spot longitude variations}
The large shift of the bright phase with respect to the eclipse
is interesting as such. RC94 noticed that since the early 
observations in 1980 the bright phase was continuously shifted
from negative longitudes to about zero longitude in 1992 and they
deduced a yearly shift of the spot longitude of 2.05\degr.
The data collected in Tab.~\ref{t:feat} imply that this 
shift might become even larger, $\Delta \chi = 2.5\degr$\,yr$^{-1}$.
Fig.~\ref{f:bripha} shows the synopsis of all the bright phase 
center (spot longitude) measurements. The phase shift is continuous
and monotonic over the last 20 years. The spot started at negative 
longitudes, i.e.~in the half-sphere away from the ballistic
accretion stream and now approaches the more typical (natural ?)
location at about 30\degr\ (see Cropper (1988) for a compilation of
accretion spot longitudes).

Spot longitude variations can be caused by changes of the mass accretion 
rate, by synchronization oscillations or by an asynchronously rotating
white dwarf. Accretion rate changes would not imply a monotonic
phase shift, they would imply a positive spot longitude at high 
accretion rate and a smaller longitude 
at low accretion rate. Since the accretion rate most probably did not 
change considerable between the \ros\ and the \xmm\ observations,
spot longitude are difficult to explain this way.

Synchronization oscillations are predicted to occur 
once a locked state between the white 
dwarf and the secondary star is reached (Campbell 1989, 
King \& Whitehurst 1991). 
So far no measurement 
could be performed in order to test the theory, the relevant 
time-scales and the amplitudes of these oscillations.
The predicted period of small oscillations about the locked state 
is $P_{\rm osc} \simeq 25$\,yr (Campbell \& Schwope 1999), 
i.e.~of the order of the time base 
covered meanwhile by the observations. 
There is no indication of a
reversal of the spot longitude migration implied by an oscillation 
scenario. We therefore tend to favor the scenario of a dis-locked white 
dwarf and thus add DP Leo to the small sub-class of asynchronous 
polars with so far four members only (Campbell \& Schwope 1999).
If our assignment is correct, 
DP Leo is different from the other systems in this sub-class 
showing a much smaller
degree of asynchronism. RC94 already estimated the deviation 
$(P_{\rm orb} - P_{\rm rot}) / P_{\rm orb} \simeq 10^{-6}$, whereas
the absolute of this quantity in the other  four is $\sim10^{-2}$.
We note that we cannot properly measure the spin period of the white 
dwarf in \dl, since the accretion spot is not fixed in the magnetic 
coordinate system of the white dwarf. 
Should the degree of asynchronism be of the order as derived here,
a fundamental re-arrangement in terms of a pole-switch must occur 
sometimes in the not too far future.

\subsection{X-ray emission from the secondary star?}\label{s:sec}

We searched the \xmm\  data for photons in the eclipse, which would be 
ascribed to the putative active secondary star. 
Omitting the first and last 10 seconds of the eclipse 
the total exposure time in eclipse investigated by us 
was 1436~s and included 6 eclipses (three cameras) in 
good time intervals. In the source-plus-background region
29 photons were registered, while in the neighboring 
background region only 18 photons were registered. 

In order to estimate the likely count rate only from the source, 
we employed a Bayesian estimate using a method described by 
Loredo (\cite{loredo}), which is applicable to a dataset with low number 
of counts having a Poisson distribution.

The most probable value of the count rate was 0.0075~cts~s$^{-1}$ taken from 
the evaluated full Bayesian probability distribution function. A Bayesian
credible region (a 'posterior bubble') 
is $0.0030-0.0120$\,cts~s$^{-1}$ in a 68\% (1$\sigma$) 
confidence interval.
The 99.73\% (3$\sigma$)
credible region gives a value $0.0000-0.0212$\,cts~s$^{-1}$,
consistent with zero. We regard our finding as uncertain marginal
detection of the secondary in X-rays.

With the nominal count rate the luminosity of the secondary is
\be
L_{\rm X} = 2.5 \times 10^{29} 
(D/400{\rm pc})^2~{\rm ergs~s}^{-1}~(0.20-7.55 \mbox{keV}).
\ee
For this estimate we used a count  to flux conversion factor 
$\sim$~1.6~$ \times 10^{-12}$
ergs~cm$^{-2}$~s$^{-1}$~cts$^{-1}$, adopted from a spectral study of one 
of the M5 type stars available in the \xmm\ Lockman Hole data 
(see also Hasinger et al \cite{hetal}).

This estimate is consistent with coronal emission of late type stars from
the solar vicinity (H\"unsch et al. \cite{huensch}).

\section{Summary and conclusion}
We have analyzed \xmm\ observations of the eclipsing polar \dl\ performed 
in the Calibration/Performance Verification phase in November 2000
in parallel with former \ros-PSPC observations.
The center of the bright phase indicates a accretion spot 
longitude of $\sim$24\degr\ at the epoch of the \xmm\ observations.
Compared to former observations, it was shifted towards later
phase, a continuation of a trend over the last 20 years. This finding
is regarded as 
indicative of an asynchronous white dwarf, although synchronization 
oscillations around an equilibrium position cannot be ruled out.
The difference between the binary period, $P_{\rm orb}$,
and the white dwarf rotation period, $P_{\rm rot}$,
is small $(P_{\rm orb} - P_{\rm rot})/ P_{\rm orb} \simeq 10^{-6}$, 4 orders
of magnitude smaller 
than for any other of the presently known four asynchronous polars. 
We have derived accurate ephemerides for the center of the 
X-ray eclipse and inferior conjunction of the secondary in \dl, 
corrected for the leap seconds introduced between 1979.9 and 2000.9
and reduced to the solar system barycenter. This can be used as 
reference for further studies of the evolution of the period of the binary 
and of the spin period of the white dwarf. 
Although we were able to correct the observed X-ray eclipse 
times to true binary phase zero through eclipse modeling, 
a direct determination of the conjunction is highly desirable.
This would allow the proper measurement of a possible spin-up or 
spin-down of the white dwarf in \dl.
The only way to achieve this in \dl\ is through 
high-speed photometry (preferentially in the ultra-violet), 
thus revealing the white dwarf. The determination of 
conjunction of the secondary star be e.g.~a spectral tracing
seems to be impossible due to the faintness of this 
low-mass star. It was not yet 
detected spectroscopically, Bailey et al.~(1993) photometrically derive 
an $R$-band magnitude of $21\fm8$. 

It is interesting to note, that only the spot longitude displays
large-scale shifts but not the latitude. A shift in latitude would result in 
a different length of the bright phase. No such effect is observed, the length
of the bright phase is always $\sim$0.57 phase units. Should an oscillation 
scenario be applicable to \dl, the non-observed latitudinal shift is an 
important extra datum for theory. The model of e.g.~King \& Whitehurst (1991)
predicts a large-amplitude out-of-plane oscillation (60\degr\ or more). 

Our detection of the secondary in X-rays is marginal, the derived 
flux and luminosity are in agreement with that of single M-stars, which rotate
much slower. However, a secure statement about the X-ray flux of the
secondary star requires a much deeper exposure.

\begin{acknowledgements}
We acknowledge constructive criticism of our referee, Dr.~W.~Priedhorsky.

Based on observations obtained with XMM-Newton, an ESA science mission 
with instruments and contributions directly funded by ESA Member 
States and the USA (NASA).

The \ros\ project is supported by the Bundesministerium f\"ur Bildung,
Wissenschaft, Forschung und Technologie (BMBF/DLR) and the Max-Planck 
Gesellschaft.

This project was supported by the 
Bundesministerium f\"ur Bildung und Forschung through 
the Deutsches Zentrum f\"ur Luft- und Raumfahrt e.V. (DLR) 
under grant number 50 OR 9706 8. 
\end{acknowledgements}

{}

\begin{thebibliography}{}

\bibitem[]{}
	Bailey J., Wickramasinghe D.T., Ferrario L., Hough J.H., 
	Cropper M., 1993, MNRAS 261, L31

\bibitem[]{}
	Biermann P., Schmidt G.D., Liebert J., et al., 1985, ApJ 293, 303

\bibitem[]{}
	Campbell C.G., 1989, MNRAS 236, 475

\bibitem[]{}
	Campbell C.G., Schwope A.D., 1999, A\&A 343, 132

\bibitem[]{}
	Caillault J.-P., Patterson J., 1990, AJ 100, 825

\bibitem[]{}
	Catal\'an M.S., Schwope A.D., Smith R.C., 1999,
        MNRAS 310, 123

\bibitem[]{}
	Chanan G.A., Middleditch J., Nelson J.E., 1976, ApJ 208, 512

\bibitem[]{}
	Cropper M., 1988, MNRAS 231, 597

\bibitem[]{}
	Cropper M., Wickramasinghe D.T., 1993, MNRAS 260, 696

\bibitem[]{}
	Cropper M., Wu K., Ramsay G., 1999, ASP Conf. Ser. 157, 325

\bibitem[1997]{chp}
	Csorg\"o, M., Horva\'ath L., 1997, Limit Theorems in 
	Change-point Analysis, New-York 

\bibitem[1999]{me} 
	Hambaryan V., Neuh\"auser R., Stelzer B., 1999, A\&A, 345, 121

\bibitem[2001]{hetal} 
	Hasinger, G., Altieri B., Arnaud, M. et al., 2001, A\&A, 365, L45

\bibitem[1999]{huensch} 
	H\"unsch M., Schmitt J. H. M. M., Sterzik M. F.,
	Voges W.G., 1999, A\&AS, 135,, 319.

\bibitem[]{}
	King A.R., Whitehurst R., 1991, MNRAS 250, 152

\bibitem[1997]{jaynes} 
	Jaynes E.T., 1997, Probability Theory: The Logic
	of Science, available at http://bayes.wustl.edu

\bibitem[1961]{jeff} 
	Jeffreys H., 1961, Theory of Probability, (3rd ed.), 
	Oxford University Press

\bibitem[1992]{loredo} 
	Loredo T.J., 1992, The Promise of Bayesian Inference for Astrophysics,
	in Statistical Challenges in Modern Astronomy, 
	ed.E.D. Feigelson and G.J. Babu (New York: Springer-Verlag),
	 pp. 275--297 (1992).

\bibitem[]{}
	Nauenberg M., 1972, ApJ 175, 417

\bibitem[]{}
	Ramsay G., 2000, MNRAS 314, 403 

\bibitem[]{}
	Pandel D., Cordova F.A., Shirey R.E., Ramsay G.,
 	Cropper M., Mason K.O., Kilkenny D., 2001, BAAS 198, 1105

\bibitem[]{}
	Ramsay G., Cropper M., Cordova F., Mason K., Much R., 
	Pandel D., Shirey R., 2001, astro-ph/0107050

\bibitem[1994]{rob_cord} 
	Robinson C.R., Cordova F.A., 1994, ApJ 437, 436

\bibitem[1998]{scarbap} 
	Scargle J. D., Bapu G. J., 1998, Point
	Processes in Astronomy: Exciting Events in the Universe, preprint,
	available at http://ccf.arc.nasa.gov/$\sim$scargle/hand$\_$www.ps

\bibitem[1998]{scar1} 
	Scargle J., 1998, ApJ, 504, 405

\bibitem[2000]{scar2} 
	Scargle J., 2000, 19th International Workshop on Bayesian 
	Inference and Maximum Entropy Methods (MaxEnt '99), 
	August 2-6, 1999, Boise State 
	University, Boise, Idaho, USA. 
	The conference proceedings will be published (Josh Rychert, Editor)

\bibitem[]{} 
	Schwarz R., Hedelt P., Rau A., Staude A., Schwope A.D., 2001, 
	ASP Conf Ser., in press

\bibitem[]{} 
	Schaaf R., Pietsch W., Biermann P., 1987, A\&A 174, 357

\bibitem[]{} 
	Schwope A.D., Schwarz R., Sirk M.M., Howell S.B., 
	2001a, A\&A, in press

\bibitem[]{} 
	Schwope A.D., Thomas H.-C., H\"afner R., Mantel K.-H., 
	2001b, A\&A, submitted

\bibitem[2001]{struder}
	Stockman H.S., Scmidt G.D., Liebert J., Holberg J.B., 1994,
	ApJ 430, 323

\bibitem[2001]{struder}
	Str\"uder L., Briel U., Dennerl K., et al., 2001, A\&A, 365, L18

\bibitem[]{}
	Warren J.K., Sirk M.M., Vallerga J.V., 1995, ApJ 445, 909

\bibitem[2001]{turner}
	Turner M. J. L., Abbey A., Arnaud M., et al. 2001, A\&A, 365, L27

\bibitem[1998]{zimm} 
	Zimmermann H. U., Boese G., Becker W., et al.,
	1998, EXSAS Users Guide, 
	\ros\ Scientific Data Center, Garching

\end{thebibliography}
\end{document}